\documentstyle[11pt]{article}
\title{\bf Quasiclassical path-integral approach to quantum mechanics
associated with a semisimple Lee algebra}
\author{ \vspace*{2mm}
{\bf E.A. Kochetov} \\
\small \it Bogoliubov Laboratory of Theoretical Physics,\\
\small \it Joint Institute for Nuclear Research, 141980 Dubna, Russia}

\begin{document}
\date{}
\maketitle
\begin{abstract}

A closed (in terms of classical data) expression for a transition
amplitude between two
generalized coherent states associated with a semisimple Lee algebra
underlying the system
is derived
for large values
of the representation highest weight,
which corresponds to the quasiclssical approximation.
Consideration is based upon a path-integral formalism adjusted
to quantization of
symplectic coherent-state manifolds that appear as one-rank coadjoint orbits.

\end{abstract}

\vspace{1cm}
{\Large \bf I. Introduction}
\vspace{1cm}

As is known, quasiclassical quantization of a classical mechanics
on a flat configuration space in terms of a Feynman path integral
results in a famous
van Vleck-Morette formula for a quantum-mechanical transition amplitude
(quantum propagator) \cite{van} which follows from an appropriate
expansion of an action about classical solutions up to terms coming from
the gaussian fluctuations and occurs in powers of the Planck constant $\hbar$,
with a classical limit corresponding to $\hbar\to 0$. In a curved
configuration space the van Vleck-Morette propagator is generalized to the
DeWitt representation \cite{dewitt1}, see also Kleinert \cite{kleinert},
and is defined entirely in terms of the classical action, trajectories
and metric.

On the other hand, a number of physically relevant models admit a natural
representation on phase spaces that appear as orbits of Lie groups in the
coadjoint representations (coadjoint orbits). This happens whenever a
Hamiltonian can be written in terms of generators of the corresponding Lie
algebra, the Hamiltonian is then said to be associated with
this algebra. Examples are obvious: spin systems with Hamiltonians being
bilinear combinations of the $SU(2)$ generators, models in non-linear
quantum optics with the relevant $SU(1,1)$ dynamical symmetry,
strongly correlated electron systems (the $t-J$ model,
the periodical Anderson model, etc.) with the underlying
$SU(2|1)$ supersymmetry.
In fact, any homogeneous symplectic manifold
that admits a connected group of isometries $G$ is locally homeomorphic
to a certain coadjoint orbit of $G$. In view of this, an appropriate
quantization of coadjoint orbits of Lie (super)groups seems to provide an
adequate basis to treat associated quantum systems.

Path integrals in quantum mechanics fall into two general categories:
phase space path integrals and configuration space ones.
The most important distinction between them is that
a time-lattice image of the phase space integral involves canonical
Liouville measure, which makes it possible to employ powerful
methods of Hamiltonian (symplectic) geometry.
Generic feature of the phase space path-integral representation
of a transition amplitude is that it implies that the corresponding
initial and final states are to be taken in {\it different}
polarizations.
Phase space path integrals including
those on coadjoint group orbits may in turn be divided into two groups
depending on whether they are written in real or complex polarizations.
A certain polarization of a phase space $M$ is required to be fixed in order
to single out, of an originally $G$-reducible Hilbert space of quantum
states, an irreducible
component ${\cal H}(M)$ which can be identified with a physical state
space \cite{kost}.

Path-integral quantization of a group action on the coadjoint orbit
in a real polarization has been considered in Refs. \cite{aleks}.
This approach incorporates Kirillov's two-form $w$ in canonical
(action-angle) variables and expresses the purely geometric part of
an action in terms of symplectic potential $\theta$, so that locally
$w=d\theta$. In practice it is more convenient however to deal with the
complex (K\"ahler) polarization corresponding to the coherent-state
representation.
Coherent states are particularly convenient since they are in one-to-one
correspondence with points of an orbit of the coadjoint representation.
Furthermore, they are the coherent states in the K\"ahler polarization
which possess {\it holomorphic} properties.

As is known, coherent states form an overcomplete basis in
${\cal H}(M)$ whenever $G$ possesses unitary irreducible square integrable
representations $U(G)$.
It then follows that
the corresponding coherent-state manifold $M$ can be viewed as a complex
{\it K\"ahler} manifold, that is, its metric, symplectic two-form, etc.
can be expressed through a single function on $M$, a K\"ahler
potential $F$. It is important that due to Berezin \cite{ber} and Onofri
\cite{onofri} the K\"ahler potential can be written down explicitly
in terms of coherent states. As a result, the path integral is entirely
determined by the K\"ahler potential ( that affects a purely geometric part
of an action)
and a classical Hamiltonian on $M$ (that determines dynamics).

Quantization of a group action on an orbit implies also that the
dominant (highest) weight of the corresponding representation comes
to play the role of the Planck constant $\hbar$, the large values of the
dominant weight corresponding to classical limit. For a spin system with the
total spin $s$ quasiclassics occurs provided inequality  $2s\gg 1$ holds,
whereas in the case of $SU(1,1)$ classical limit corresponds to large values
of occupation numbers \cite{perel}. In this regard, quasiclassical
quantization by the coherent-state path integral may provide some
nontrivial asymptotic representations of physical relevance.
A conventional configuration space path integral could hardly be used for
this purpose, since certain constraints are to be additionally imposed
to fix a representation, which usually results in a severe technical
problem.

The purpose of the present paper is to derive a closed representation
of the coherent-state propagator (matrix element of an evolution operator)
for large values of the dominant weight $l$ that specifies the
representation $({\cal H}(M),U(G))$. To this end, the corresponding
quantization by the coherent-state path integral turns out to be
convenient since it properly takes into account the underlying symmetry.
The total action turns out to be linear in
$l$, which makes it quite reasonable
to apply the stationary-phase approximation in evaluating the integral
as $l\to \infty$.

As we are interested in the path-integral representation for a transition
amplitude, we should care about the so-called
boundary term in the continuum action, the latter disappearing in the path
integral for {\it a partition function}.
This term is also expressible via the
K\"ahler potential as has been shown for ordinary coherent states
\cite{koch1} and supercoherent states \cite{koch2} and is of
crucial importance in deriving a correct quasiclassical
($l\to \infty$) coherent-state propagator. Otherwise some
difficulties arise, e.g., the so-called overspecification problem
\cite{klaud1} and contradictions with the path-integral generalization
of the Duistermaat-Heckman (DH) theorem appear. This was just the case
in some earlier attempts to derive the quasiclassical $SU(2)$ propagator
\cite{kurat}, whereas the correct expression for the $SU(2)$ case has recently
been obtained \cite{koch3}.

In the present paper we expound an analogous derivation
for a real semisimple Lie group $G$ possessing square integrable
representations.
As is known, both compact and noncompact groups with the discrete series
representations fall into this category,
the results we obtain are also equally well applied
to the widely used Heisenberg-Weyl coherent states parametrized by points
of a complex plane $C$.

The plan of the paper is as follows. In Section II, we present a general
coherent-state path integral for a transition amplitude between two
coherent states. Section III comprises some preliminaries and
a brief review of earlier results
on the quasiclassical evaluation of quantum-mechanical propagators.
Section IV constitutes the main result, a derivation of the
closed quasiclassical formula for the coherent-state
propagator.
Relations between various quantization schemes are discussed in Section V.
Few examples are gathered in Section VI to illustrate
possible applications of the general result.
A short summary concludes the paper in Section VII.

\vspace{1cm}
{\Large \bf II. Coherent-state path integral}
\vspace{1cm}

As is known, Perelomov's coherent states for a semisimple group $G$
are points
of an orbit of a unitary irreducible representation of $G$
in an abstract Hilbert space ${\cal H}$ \cite{perel}.
By choosing an initial state $|0\rangle$ in ${\cal H}$,
called the fiducial state the vectors of the corresponding $G$ orbit
are parametrized by points of a homogeneous space $M=G/G_0$, where $G_0$
is the isotropy subgroup of $|0\rangle$. In the following we will be
interested in the case where $|0\rangle$ appears as a dominant weight vector
(highest weight vector up to the Weyl transformation), which correspons to
the quantization in the K\"ahler (holomorphic) polarization.
It then follows that
a factor space
$G/G_0$ appears as a K\"ahler manifold, the K\"ahler potential
being directly expressible in terms of the coherent states as follows.
Given a coherent state $|z\rangle$
where $z$
belongs to $G/G_0$,
we define (locally)
\begin{eqnarray}
F(\bar z_1,z_2)=\log\frac{\langle z_1|z_2\rangle}
{\langle z_1|0\rangle\langle 0|z_2\rangle}
\label{eq:2.1}\end{eqnarray}
which can be viewed as an analytic continuation of the real valued
function
\begin{eqnarray}
F(\bar z,z)=\log|\langle 0|z\rangle|^{-2}.
\label{eq:2.2}\end{eqnarray}
The latter is called the K\"ahler potential and was introduced in this way
by Berezin \cite{ber} and Onofri \cite{onofri}.
This function incorporates geometry
of the underlying phase space and plays a crucial role in the following.

The phase space $G/G_0$ can be equipped with an invariant
supersymplectic 2-form $w$,
\begin{eqnarray}
w\equiv-i\delta\bar\delta F(\bar z,z),
\label{eq:2.3}\end{eqnarray}
where $\delta =dz\otimes\partial/\partial z$
and
$\bar\delta =d\bar z\otimes\partial/\partial\bar z$
such that the exterior derivative
$d\equiv \delta+\bar\delta$. A straightforward calculation shows that $w$
is closed, i.e., $dw=0$, which means that $G/G_0$ is a symplectic
manifold, in other words, it may serve as a classical phase space.
In terms of $F$, metric and $G$-invariant Liouville measure read
\begin{eqnarray}
ds^2=g\,dzd\bar z= \partial^2_{\bar zz}F\,dzd\bar z
\label{eq:2.4}\end{eqnarray}
\begin{eqnarray}
d\mu\sim \,g\,\frac{dzd\bar z}{2\pi i},
\label{eq:2.5}\end{eqnarray}
which yields
\begin{equation}
\int|z\rangle\langle z|d\mu=I,
\label{eq:2.6}\end{equation}
Resolution of unity (\ref{eq:2.6}) enters into a path integral as
a basic ingredient.

Consider the quantum propagator in the $z$-representation
\begin{equation}
\langle z_F|T\,\exp-\frac{i}{\hbar}\int\limits_{0}^{\tau}H(s)ds|z_I\rangle
\equiv
{\cal P}({\bar z}_F,z_I,;\tau)\,,
\label{eq:2.7}
\end{equation}
which represents
Berezin's covariant symbol of the evolution operator, the Hamiltonian $H(s)$
being a polynomial function of the $G$ generators with time-dependent
coefficients.
In Eq. (\ref{eq:2.7}) $T$ denotes the time-ordering
symbol. It is necessary to include, because the operator $H(s)$
in the exponent
does not commute for different values of $s$.

In order to express the transition amplitude by a path integral, we divide
the time interval into $N$ small intervals: $\epsilon=\tau/N$ with
$N\to\infty$. Let us define
$$s_k=\epsilon k,\quad z_k=z(s_k),
\quad 0\le k\le N.$$
With the aid of the time discretization together with relation (\ref{eq:2.6})
the propagator can be written up to first order in $\epsilon$ in the form
\begin{eqnarray}
{\cal P}&=&\lim_{N\to\infty} \int_{z_0=z_I}^
{\bar z_N=\bar z_F} \prod_{k=1}^{N-1}d\mu_k
\langle z_F|z_{N-1}\rangle\langle z_1|z_I\rangle
\nonumber\\
&&\times \prod_{k=2}^{N-1}\langle z_k|z_{k-1}\rangle
\exp\{-i\epsilon \sum_{k=1}^{N}H^{cl}
(\bar z_k,z_{k-1})\},
\label{eq:2.8}\end{eqnarray}
where
$$H^{cl}(\bar z_k,z_{k-1};s_k)=
\frac{\langle z_k|H(s_k)
|z_{k-1}\rangle}
{\langle z_k|z_{k-1}\rangle}.$$
The variables $z_N$ and $\bar z_0$
do not enter into Eq. (\ref{eq:2.8}) at all.
In other words, the Euler-Lagrange
equations are accompanied by the
boundary conditions $z_0=z_I$ and
$\bar z_N=\bar z_F$, respectively.
The term
$$\langle z_F|z_{N-1}\rangle\langle z_1
|z_I\rangle$$
gives rise to the continuum boundary term to be discussed below.

Before proceeding further, the following remark on representation
(\ref{eq:2.8}) applies.
As is known, if a phase space is compact, it cannot be covered by a single
chart.
On the other hand, every integral in Eq. (\ref{eq:2.8}) is written
in one local chart. The way out is that the phase space $M$ is
$G$-homogeneous (the group $G$ acts on $M$ through holomorphic isometries),
so that a full set of local charts is generated by actions of $G$:
any two charts are locally related by $z\to gz$, for some $g\in G$.
Strictly speaking, any (ordinary) integral in Eq. (\ref{eq:2.8})
should have been written by the definition
\begin{eqnarray}
\int_{M}d\mu\,(\cdots)=\sum_i\int_{V_i}d\mu\,(\cdots)\,\epsilon_i,
\label{eq:2.88}\end{eqnarray}
where $\{\epsilon_i\}$ is a partition of unity subordinate to the covering
$\{V_i\}$. Since each local chart covers $M$ except for a set of measure $0$
(with respect to $d\mu$), one may, by employing appropriate $G$-shifts
of variables on the lhs of (\ref{eq:2.88}) and making use of
the $\epsilon$-defining equation $\sum_i\epsilon_i=1,$ restrict
the integration to the single local chart $\cong C$.

In the continuum limit Eq. (\ref{eq:2.8}) takes on the form
\begin{equation}
{\cal P}=
\int\limits_{z(0)=z_I}^{\bar z(\tau)=\bar z_F}
D\mu(z)\exp\Phi.
\label{eq:2.9}
\end{equation}
The total action $\Phi$ includes the boundary term $\Gamma$:
$$\Phi=S+\Gamma,$$
where
\begin{equation}
S=-\frac{1}{2}\int\limits_{0}^{\tau} \left(\dot z\frac{\partial F}
{\partial z} -{\dot{\bar z}}\frac{\partial F}{\partial\bar z}
\right)ds
-\frac{i}{\hbar}\int\limits_{0}^{\tau}
H^{cl}(\bar z,z)ds,
\label{eq:2.10}
\end{equation}
\begin{eqnarray}
\Gamma=
\frac{1}{2}\left[F(\bar z_F,z(\tau))+F(\bar z(0),
z_I)
-F(\bar z_F,z_F)
-F(\bar z_I,z_I)\right].
\label{eq:2.11}
\end{eqnarray}
These equations coincide up to an obvious change in the
notation with those of Ref. \cite{koch1}.

Continuum representation (\ref{eq:2.11}) originates from a specific
discontinuity of paths $z(s)$ and $\bar z(s)$ at the relevant end points.
For example, let us introduce
$\Delta_k(\epsilon)\equiv z_k-z_{k-1}.$
Then one has $\lim_{\epsilon\to 0}\Delta_k(\epsilon)=0$ for all $k$,
except that
$\lim_{\epsilon\to 0}\Delta_N(\epsilon)\neq 0$, since $z_N=z_F$ and $z_F$
is an arbitrary complex number. However, instead of explicitly writing out
corresponding shifts of the arguments, it is more convenient to
consider variables $\bar z(s)$ and $z(s)$ to be
independent. Formally, this amounts to saying
that the initial $|z_I\rangle$ and final $\langle z_F|$
configurations are in different polarizations \cite{simms}, which
necessitates the appearance of the boundary term.
For example, consider a classical system specified by
the Hamiltonian function $h^{cl}$
with initial and final configurations being taken
in the polarizations generated by
$\partial/\partial q$ and $\partial/\partial p$,
respectively:
\begin{eqnarray}
\dot q&=&\frac{\partial h^{cl}(q,p)}{\partial p},\quad q(\tau)=q_F
\nonumber\\
\dot p&=&-\frac{\partial h^{cl}(q,p)}{\partial q},\quad p(0)=p_I.
\end{eqnarray}
These equations follow from the Hamilton principle of stationary
action $\delta \phi =0$,
where
$$\phi=i\int_{0}^{\tau}[p\dot q-h^{cl}]ds-ip_I[q_F-q(0)].$$

To specify the classical system associated with $H$, one needs classical
equations of motion. The latter follow from the Hamilton principle
$\delta\Phi=0$, which yields
\begin{eqnarray}
&&{\dot{\bar z}}=i{\hbar}^{-1}(\partial^2_{\bar zz}F)^{-1}
\partial_z H^{cl}\,\,\,\,\,\,\,
\bar z(\tau)=\bar z_F\,,\nonumber \\
&&\dot z=-i{\hbar}^{-1}(\partial^2_{\bar zz}F)^{-1}
\partial_{\bar z} H^{cl}\,\,\,\,\,\,\,
z(0)= z_I\,.
\label{eq:2.12}
\end{eqnarray}
One sees from (\ref{eq:2.12}) that equations of motion are
correctly specified by boundary conditions and define a canonical
phase flow associated with $H^{cl}$.

In view of a rather complicated form of the coherent-state path integral,
it would be desirable to get simple sufficient criteria for the
stationary-phase approximation to be exact. These are provided by the
path-integral generalization of the DH theorem which states (omitting
some subtleties) that
the WKB approximation is exact, provided the Hamiltonian flow leaves
a metric of the underlying phase space invariant, that is
\begin{eqnarray}
{\cal L}_{X_H}\,g=0,
\label{eq:lie}\end{eqnarray}
where ${\cal L}_{X_H}$ stands for a Lie derivative along a
Hamiltonian vector field that generates the flow.

To formally apply the DH theorem a kinetic term in
an action is required to be of the form $i\int \theta$,
where the symplectic 1-form $\theta$ determines $w$ by
$d\theta=w$ \cite{niemi}.
This is pursued for the representation
(\ref{eq:2.9}-\ref{eq:2.11}) as follows.
Let us define
\begin{eqnarray}
\theta=\frac{i}{2}\left[\delta F(\bar z,z)
-\delta F(\bar z_F,z)
-\bar\delta F(\bar z,z)
+\bar\delta F(\bar z,z_I)\right].
\label{eq:2.13}\end{eqnarray}
By the very construction,
$d\theta=w.$
We recall that $d=\delta+\bar\delta$ and $\delta^2=\bar\delta^2=
\delta\bar\delta+\bar\delta\delta=0$.
A straightforward computation yields $$ S+\Gamma= i\int\theta
-\frac{i}{\hbar}\int_{0}^{\tau}H^{cl}ds
-\frac{1}{2}\left[F(\bar z_F,z_F)
+F(\bar z_I,z_I)-2F(\bar z_F,z_I)\right]$$
$$\equiv i\int\theta -\frac{i}{\hbar}
\int_{0}^{\tau}H^{cl}ds+\log\langle z_F,|z_I\rangle,$$
which results in a desired representation,
$$\frac{\langle z_F|T\,\exp-\frac{i}{\hbar}\int\limits_{0}^{\tau}H(s)ds
|z_I\rangle}{\langle z_F|z_I\rangle}=
\int\limits_{z(0)=z_I}^{\bar z(\tau)=\bar z_F}
D\mu(z)\exp\left[i\int\theta
-\frac{i}{\hbar}\int_{0}^{\tau}H^{cl}ds\right].$$

To avoid a possible confusion, we conclude this section with the following
remark. The quantization by means of a path integral depicted above
provides an example of the so-called
quantization-versus-classical-limit procedures.
We start with the quantum Hamiltonian $H$, evaluate its classical limit
through the associated coherent states and then quantize the obtained
classical system $(H^{cl},G/G_0,w)$ by using the path integral on $G/G_0$.
At first sight that may seem strange, but the point is that
we start with {\it an abstract} representation of $H$ and end up with
{\it the explicit} one. A few examples to illustrate this situation
will be given in section VI.

\vspace{1cm}
{\Large \bf III. Quasiclassical approximation: preliminaries}
\vspace{1cm}

Let $G$ be a compact simple Lie group. For any unitary irreducible
representation $U^{\vec l}(G)$, its highest weight $\vec l$ is given
by a sum of the fundamental weights $\vec\omega^j$
with the non-negative integer coefficients
$$\vec l=\sum_{j=1}^rl^j\vec\omega^j,$$
where $r$ stands for the rank of a Lie algebra of $G$.
It then can be shown that
\begin{eqnarray}
F^{\vec l}=\sum_{j=1}^rl^j F^j ,
\label{eq:3.1}\end{eqnarray}
where $\{F^j\}$ represent the fundamental K\"ahler potentials.
The same dependence on $l^j$ holds for the covariant symbols
(coherent state expectation values)
of the basic elements of the Lie algebra of $G$ \cite{marinov}.
In the sequel, for the sake of simplicity, we will be solely concerned
with the case when only one term in the series contributes to (\ref{eq:3.1}),
which corresponds to group orbits
of $rank=1$. Given a group representation $U^{\vec l}(G)$,
the number of the non-zero nodes $l_j\neq 0$ in the corresponding
Dynkin graph may be called {\it the rank} of $M$ \cite{marinov}.
If all $l_j >0$ the corresponding orbit
is called a non-degenerate one.
This requirement by no means
restricts oneself to the groups $G=SU(2)$ and $SU(1,1)$
and corresponding homogeneous spaces $SU(2)/S(U(1)\otimes U(1))$
and $SU(1,1)/S(U(1)\otimes U(1))$, for
there exist one-rank manifolds (degenerate orbits of higher rank groups)
with complex dimensions $dim_c M>1$.

In physics the symplectic form $w$ has the same units as an angular momentum
$$[w]=[kg\, m^2\, sec^{-1}]=[\hbar].$$ Since we assume the coordinates
$z, \bar z$ to be dimensionless, the form
$w$ in Eq. (\ref{eq:2.3}) is implicitly
understood to be measured in units of $\hbar$.
It is convenient to introduce a new parameter
$$\lambda=\hbar l,$$
that represents a physical quantity whereas $\hbar$ represents the
quantum mechanical yard stick with which to measure $\lambda$ \cite{tuynman}.
For instance, in the case $(M=S^2,l\in N)\,\lambda$
represents the intrinsic angular momentum (spin).
It is the parameter $\lambda$ that enters into measurable
physical quantities, e.g., energy. In order to keep them fixed in the limit
$l\to\infty$, one should simultaneously imply that $\hbar\to 0$.
\footnote{The limit $\hbar\to 0$ means a passage from systems of units
well adjusted for describing quantum objects to those
which are more suitable for the classical ones.} This notice explains why the
large highest weight $l$ corresponds to a quasiclassical region.

In the classical theory, the Lie algebra is represented by covariant symbols
or momentum maps which are functions on $M$, with the Lie product being
the Poisson brackets given by the K\"ahler structure. In the limit
$l\to\infty\,(\hbar\to 0)$ the algebra of operators (observables)
corresponding to the Lie generators of $G$ reduces to the Poisson algebra
of functions (momentum maps) on $M$. To put another way,
in the classical limit orbits in the coadjoint representations of $G$ emerge,
different representations giving rise to different orbits.
A method of explicit obtaining classical phase spaces (group orbits)
for the case of $G=SO(3)$ has been worked out by Lieb \cite{lieb}
and later on generalized to compact simple groups by Simon \cite{simon}.

Let $\{L_{\alpha}\}$ denote a set of the generators of $G$. Consider the
Hamiltonian
\begin{eqnarray}
H=\sum_{\alpha=1}^{dim G}\hbar\omega^{(1)}_{\alpha}L_{\alpha}+
\sum_{\alpha,\beta=1}^{dim G}\hbar\omega^{(2)}_{\alpha\beta}
\phi^{(2)}_{\alpha\beta}(l)L_{\alpha}L_{\beta}+\ldots
\label{eq:3.2}\end{eqnarray}
where $\omega^{(1)}_{\alpha},\omega^{(2)}_{\alpha\beta},\ldots$
are some frequencies
that may explicitly depend on time
and functions $\phi^{(2)}_{\alpha\beta}(l),\ldots$ are choosen
to ensure that $H^{cl}$ linearly depends on $\lambda$. In view of
the aforementioned property of momentum maps, this automatically holds for
the first term. \footnote{ Numerous spin-spin lattice interactions
fall into this category.
A simple
example of function $\phi^{(2)}$ is provided by
$$H\sim\frac{1}{2j-1}(J_{+}^2+J_{-}^2),\quad \langle H\rangle\sim2j=l,
\quad \phi^{(2)}\sim \frac{1}{2j-1},\quad j>1/2.$$
For the notation herein see sect. VI.} From Eq. (\ref{eq:3.1}) it then
follows that all dependence on $l$ is isolated in a single factor of $(l)$
multiplying the total action $\Phi$,
which justifies the application of the stationary phase
approximation to the path-integral
propagator (\ref{eq:2.9}).

To conclude this section, a few remarks about already known quasiclassical
formulas for quantum-mechanical propagators are to be made.
The main result in this respect is due to DeWitt and leads to the short
time approximation to a transition amplitude in a configuration space
\cite{dewitt1}
\begin{eqnarray}
\langle q''|e^{-\frac{i}{\hbar}(t''-t')H_{+}}|q'\rangle=
(\frac{i}{2\pi\hbar})^{r/2}g''^{-1/4}D^{1/2}
(q''t''|q't')g'^{-1/4}\exp\{\frac{i}{\hbar}S(q''t''|q't')\}
+{\cal O}(\Delta t^2),
\label{eq:3.3}\end{eqnarray}
where $g'=\det|g_{ij}(q')|$, etc., $q$ stands for a point of an $r$-dimensional
Riemannian space of metric $g_{ij}$;
$$S=\int_{t'}^{t''}\left(\frac{1}{2}g_{ij}\dot q^i\dot q^j-v(q)\right)dt
\equiv \int_{t'}^{t''} {\cal L}dt$$
is the classical action along extremal with $q(t')=q',q(t'')=q''$.
The Hamiltonian $H_{+}$ differs from that corresponding to ${\cal L}$
by the term that involves a scalar curvature $R$:
$$H_{+}=H+\kappa\hbar^2 R,$$
where $\kappa$ is a numerical constant.
The van Vleck determinant
$$D=\det\left[\frac{\partial^2S}{\partial q''^j\partial q'^i}\right].$$
In the flat case
$(g_{ij}=\delta_{ij})$ Eq. (\ref{eq:3.3}) goes over to the well-known
van Vleck-Morette formula.
The matrix element on the l.h.s. of Eq. (\ref{eq:3.3}) can be written down
as {\it a configuration space} path integral:
\begin{eqnarray}
\langle q''|e^{-\frac{i}{\hbar}(t''-t')H_{+}}|q'\rangle=
\int\,D\mu(q)\exp\int_{t'}^{t''}{\cal L}_{-}dt,
\label{eq:3.4}\end{eqnarray}
where ${\cal L}_{-}={\cal L}+\kappa\hbar^2 R$.
As has been shown by Kleinert, the non-classical term in the action
proportional to $\hbar^2$ disappears, provided one treats the differences
$\Delta q^{\mu}$ and differentials $dq^{\mu}$ symmetrically, when transforming
the integral measure in the cartesian path integral into the general
metric-affine space \cite{kleinert}. This results in a correct energy
spectrum of a free particle on surfaces of spheres and on group
manifolds obtained from a path integral with a purely classical action in the
exponential.

If one neglects the $R$-depenent term $\sim\hbar^2$,
Eq. (\ref{eq:3.3}) can be regarded as a quasiclassical $(\hbar\to 0)$
approximation to the transition amplitude.
An inconvenient point in this approach is, however, that there is
no simple sufficient
criteria for the quasiclassical expression (\ref{eq:3.3}) to be exact.
Since the configuration space path integral in contrast to the
phase space one does not involve any Liouville
measure (in a time-lattice discretization), the DH theorem cannot be
applied in this case. Yet some "experimental" observation can be made.
As has been noted by Schulman \cite{schulman1} and DeWitt \cite{dewitt2},
the configuration space path integral is WKB exact if the expression
for the {\it finite} time propagator concides with that for the short time.
As is shown by Dowker \cite{dowker} and
Marinov and Terentyev \cite{marinov2},
this occurs for a free particle propagating on a group manifold.
For further development of these ideas the
recent papers by Inomata
\cite{inom} and Junker \cite{junker} could be referred to.

The coherent-state path-integral formalism turns out to be more convenient
since a powerful machinery of canonical transformations can be employed
and a hidden supersymmetry of an action can be revealed, which leads to
the path integral generalization of the DH theorem that provides
{\it universal} simple criteria (\ref{eq:lie})
for the quasiclassics to be exact
\cite{niemi}. Besides that, this method incorporates the
underlying symmetry of the problem under consideration, which makes it
possible to look for an asymptotic behavior with respect to the
representation indices (eigenvalues of Kasimir's operators).

It was Klauder \cite{klaud1} as well as Klauder and Daubechies \cite{daubech}
who first suggested to use a system of the type (\ref{eq:2.12}) to derive
the semiclassical approximation for the coherent-state path integral.
Identity of the direct semiclassical evaluation of the
coherent-state propagator on a complex plane to the stationary phase
approximation to the corresponding path integral
except for the pre-exponential factors was proved by Weissman
\cite{weiss2}. In the important work by Yaffe \cite{yaff}
a general method for finding classical limits in certain quantum theories
was developed. This approach is naturally based upon coherent states
associated with a symmetry group and is used to explicitly
construct a classical phase space, a corresponding coadjoint orbit.
Non of the symmetry groups considered in this paper are semisimple,
which makes it necessary to distinguish between adjoint and coadjoint
orbits. Cadavid and Nakashima \cite{cad} studied the coherent-state
path integral for semisimple Lie algebras, coherent states being
sections of a holomorphic line bundle over $G/G_0$.
The semiclassical approximation
of the quantum evolution operator via coherent states associated with
quantized closed curves on the $SU(2)$ orbits was obtained by
Karasev and Kozlov \cite{kar1}. This method was further extended
to semisimple Lie algebras \cite{kar2} and general K\"ahler phase
spaces \cite{kar3}.

\vspace{1cm}
{\Large \bf IV. Quasiclassical approximation: coherent-state propagator}
\vspace{1cm}

In this section we present a derivation of the quasiclassical
coherent-state propagator ${\cal P}^{qc}$ by applying the stationary
phase approximation to the path integral (\ref{eq:2.9}).

We are looking for the representation
\begin{eqnarray}
{\cal P}=e^{l\,(\cdots)}\,[(\cdots)+ o(1)],\quad l\to\infty,
\label{eq:pr}\end{eqnarray}
where $(\cdots)$ stands for $l$-independent functions on a phase space.
The quasiclassical propagator is then defined by the leading term of
(\ref{eq:pr})
\begin{equation}
{\cal P}^{qc}= e^{l\,(\ldots)}\,(\cdots).
\label{eq:qcpr}\end{equation}

We first rewrite (\ref{eq:2.9}) to explicitly allow for the normalization:
\begin{equation}
{\cal P}= \langle z_F|z_I\rangle\,
\frac{\int D\mu(z)\exp\Phi}{\int D\mu(z)\exp\Phi_0},
\quad \Phi_0\equiv \Phi\mid_{H=0}.
\label{eq:a}
\end{equation}
In order to lift the measure weight factor $\partial^2_{\bar zz}F$ in an
exponential we make use of a trick that consists in the integration
over auxiliary
anticommuting fields $\bar\xi(t)$ and $\xi(t)$ (see, e.g.,
\cite{niemi}):
\begin{equation}
{\cal P}= \langle z_F|z_I\rangle\,
\frac{\int D\bar zDzD\bar\xi D\xi\exp\,[\Phi+\int \bar\xi(t)
(\partial^2_{\bar zz}F)\xi(t) dt]}
{\int D\bar zDzD\bar\xi D\xi\exp\,[\Phi_0+\int \bar\xi(t)
(\partial^2_{\bar zz}F)\xi(t) dt]}.
\label{eq:b}
\end{equation}

The quasiclassical $(l\to\infty)$ motion
is described by the approximation
\begin{equation}
\Psi\equiv \Phi+\int \bar\xi(\partial^2_{\bar zz}F)\xi dt=
\Psi\mid_c + \frac{1}{2}\delta^2\Psi\mid_c+...\simeq\Psi\mid_c+
\frac{1}{2}\delta^2\Psi\mid_c,
\quad\delta \Psi\mid_c=0
\label{eq:4.1}
\end{equation}
and
\begin{equation}
\Psi_0\equiv \Phi_0+\int \bar\xi(\partial^2_{\bar zz}F)\xi dt=
\Psi_0\mid_c + \frac{1}{2}\delta^2\Psi_0\mid_c,
\quad\delta \Psi_0\mid_c=0
\label{eq:4.01}
\end{equation}
with the boundary
conditions $z\left(0\right)=z_I$ and $\bar z \left(\tau\right)=\bar z_F.$
The subscript $"c"$ denotes a value along the extremals
(\ref{eq:2.12}).

To proceed further, we introduce variations
$$\delta z\equiv \eta=z-z_c,\quad
\delta\bar z\equiv\bar\eta=\bar z-\bar z_c,$$
which satisfy $$ \eta(0)=0\,,\,\,\,\,\,\, \bar\eta(\tau)=0.$$
It is clear that $\bar\xi\mid_c=\xi\mid_c=0$ and in view of (\ref{eq:2.1})
$\exp\Phi_0\mid_c=\langle z_F|z_I\rangle$. Bearing this in mind we insert
expansions (\ref{eq:4.1},\ref{eq:4.01}) into Eq. (\ref{eq:b}) perform
integrals over $\delta\bar\xi,\delta\xi$ coming from $\delta^2\Psi$ and
$\delta^2\Psi_0$ that cancel each other \footnote{The nontrivial measure
$D\mu$ contributes to ${\cal P}$ in higher orders} and finally arrive at
\begin{equation}
{\cal P}^{qc}(\bar z_F,z_I;\tau)={\cal P}_{red}\,\exp\Phi_c,
\label{eq:4.2}
\end{equation}
where the reduced propagator is given by
\begin{eqnarray}
{\cal P}_{red}=&&\int D\eta D\bar\eta \exp\left\{\frac{1}{2}\int_{0}^{\tau}
(\dot{\bar\eta}\eta-\dot\eta\bar\eta)ds -\frac{i}{2}\int_{0}^{\tau}
(\eta^2A+{\bar\eta}^2C+2\bar\eta \eta B)ds\right\} \nonumber \\
=&&\left(\frac{Det K}{Det K_0}\right)^{-1/2},
\label{eq:4.3}
\end{eqnarray}
with
$$K=\left(\begin{array}{cc}-iA(s)& -iB(s)+\partial _s\\
-iB(s)-\partial _s& -iC(s)\end{array}\right) \quad\mbox{ and } \quad
K_0=\left(\begin{array}{cc}0& \partial _s\\
-\partial _s &0 \end{array}\right).$$
The functions
  $$A={\hbar}^{-1}\partial_z\left[(\partial^2_{\bar zz}F)^{-1}
\partial_zH\right]\mid_c,
\quad C={\hbar}^{-1}\partial_{\bar z}
\left[(\partial^2_{\bar zz}F)^{-1}
\partial_{\bar z}H\right]\mid_c,$$
$$B=\frac{1}{2\hbar}\partial_z
\left[(\partial^2_{\bar zz}F)^{-1}
\partial_{\bar z}H\right]\mid_c +
\frac{1}{2\hbar}\partial_{\bar z}
\left[(\partial^2_{\bar zz}F)^{-1}
\partial_ zH\right]\mid_c $$
are calculated with the aid of the Euler-Lagrange equations
(\ref{eq:2.12}).
We recall that $\eta$ and $\bar\eta$ are considered to be independent.

Our aim now is to express (\ref{eq:4.3}) in terms of the classical orbitals.
First it is convenient to eliminate the $B$-term from
$\delta^2\Phi_c(\bar\eta,\eta)$ in (\ref{eq:4.3}), which can be achieved
by the change
$$\eta\to\eta\exp(-i\int_{0}^{s}B(t)dt).$$
The result is
\begin{eqnarray}
{\cal P}_{red}=\int D\eta D\bar\eta\exp(\frac{1}{2}\delta^2\tilde{\Phi_c}
(\bar\eta,\eta)),
\label{eq:4.4}\end{eqnarray}
where
\begin{eqnarray}
\delta^2\tilde{\Phi_c}(\bar\eta,\eta)=
\int_{0}^{\tau}(\dot{\bar\eta}\eta-\dot\eta\bar\eta)ds -
i\int_{0}^{\tau}
(\eta^2\tilde A+{\bar\eta}^2\tilde C)ds)
\label{eq:4.5}\end{eqnarray}
and
$$\tilde A=A\exp(-2i\int_{0}^{s}Bdt),\quad
\tilde C=C\exp(2i\int_{0}^{s}Bdt).$$
The Euler-Lagrange equations for the functional (\ref{eq:4.5}) known as
the Jacobi equations read
\begin{eqnarray}
\tilde K \Upsilon=0,
\label{eq:4.6}\end{eqnarray}
where
$$\tilde K=\left(\begin{array}{cc}-i\tilde A(s)& \partial _s\\
-\partial _s& -i\tilde C(s)\end{array}\right) \quad\mbox{ and } \quad
\Upsilon=
\left(\begin{array}
{cc}\eta\\ \bar\eta\end{array}\right),\quad \eta(0)=0,\quad
\bar\eta(\tau)=0.$$
It is clear that
\begin{eqnarray}
{\cal P}_{red}=\left(\frac{Det\tilde K}{Det K_0}\right)^{-1/2}.
\label{eq:4.7}\end{eqnarray}
Let us denote
$$f(\lambda)=Det K_0^{-1}(\tilde K-\lambda),$$
where $\lambda$ is a spectral parameter (not to be confused with the
quasiclassical parameter). It then follows that
\begin{eqnarray}
\partial_{\lambda}\log f=-tr(\tilde K-\lambda)^{-1}.
\label{eq:4.8}\end{eqnarray}
To construct the Green function $(\tilde K-\lambda)^{-1}$, consider
the system of two equations
\begin{equation}
(\tilde K-\lambda)\Phi=0,\quad
\Phi=\left(\begin{array}
{cc}\phi\\ \bar{\phi}\end{array}\right),\quad \phi(0)=0\,
\label{eq:4.9}\end{equation}
and
\begin{equation}
(\tilde K-\lambda)\Psi=0,\quad
\Psi=\left(\begin{array}
{cc}\psi\\ \bar\psi\end{array}\right),\quad \bar\psi(\tau)=0.
\label{eq:4.10}
\end{equation}
With the aid of (\ref{eq:4.9},\ref{eq:4.10}) the Green function
is constructed to be
\begin{eqnarray}
(\tilde K-\lambda)^{-1}(t,t')&=&\frac{\theta(t-t')}{w}
\left[\begin{array}{cc}\phi(t')\psi(t)   & \bar\phi(t')\psi(t)   \\
\phi(t')\bar\psi(t)  & \bar\phi(t')\bar\psi(t)  \end{array}\right]
\nonumber\\
&+& \frac{\theta(t'-t)}{w}
\left[\begin{array}{cc}\phi(t)\psi(t')   & \phi(t)\bar\psi(t')   \\
\bar\phi(t)\psi(t')  & \bar\phi(t)\bar\psi(t')  \end{array}\right].
\label{eq:4.11}\end{eqnarray}
In the case when the functions $\Phi$ and $\Psi$ are linear-independent, i.e.,
$\Phi$ does not coincide with an eigenfunction of $\tilde K$, the Wronskian
$$w=\left|\begin{array}{cc}\phi(s) & \bar\phi(s) \\
\psi(s) & \bar\psi(s)  \end{array}\right|=\phi\bar\psi-\bar\phi\psi$$
does not depend on $s$ \cite{levitan}.
In view of (\ref{eq:4.9}, \ref{eq:4.10})
\begin{eqnarray}
w(0)=-\bar\phi(0)\psi(0)=-\bar\phi(\tau)\psi(\tau)=w(\tau)
\label{eq:4.12}\end{eqnarray} From Eq.(\ref{eq:4.11}) it immediately
follows that
\begin{eqnarray}
tr(\tilde K-\lambda)^{-1}=\frac{1}{\omega}\int_{0}^{\tau}(\phi\psi+
\bar\phi\bar\psi)dt.
\label{eq:4.13}\end{eqnarray}
This integral can be taken by using the following notice. Instead of
system (\ref{eq:4.9}, \ref{eq:4.10}) let us consider a slightly changed one:
$$(\tilde K-\lambda')\Phi=0, \quad(\tilde K-\lambda)\Psi=0 $$
After little algebra one finds
$$(\lambda'-\lambda)(\phi\psi+\bar\phi\bar\psi)=[\dot{\bar\phi}\psi-
\dot{\bar\psi}\phi+{\dot\psi}{\bar\phi}-
{\dot\phi}{\bar\psi}]\mid_{\lambda'=\lambda}$$
$$+(\lambda'-\lambda)\left[\partial_{\lambda}\dot{\bar\phi}\psi -
\dot{\bar\psi}\partial_{\lambda}\phi+{\dot\psi}\partial_{\lambda}\bar\psi
-{\bar\psi}\partial_{\lambda}\dot\phi\right]\mid_{\lambda'=\lambda}
+{\cal O}((\lambda'-\lambda)^2).$$
The first bracket on the rhs equals zero since
$$\frac{d}{ds}\,w=0,$$
and one finally ends up with
$$\phi\psi+\bar\phi\bar\psi=\frac{d}{dt}(\psi\partial_{\lambda}\bar\phi-
\bar\psi\partial_{\lambda}\phi),$$
which yields
$$tr(\tilde K-\lambda)^{-1}=-\frac{\psi(\tau)\partial_{\lambda}
\bar\phi(\tau)}{\bar\phi(\tau)\psi(\tau)}+
\frac{\psi(0)\partial_{\lambda}
\bar\phi(0)}{\bar\phi(0)\psi(0)}=-\partial_{\lambda}\log
\frac{\bar\phi(\tau;\lambda)}{\bar\phi(0;\lambda)}.$$
With the help of this one finally gets
\begin{eqnarray}
Det\,\tilde K/K_0=\frac{\bar\phi(\tau;\lambda=0)}{\bar\phi(0;\lambda=0)}
=\frac{\psi(0;\lambda=0)}{\psi(\tau;\lambda=0)}.
\label{eq:4.14}\end{eqnarray}
In the last line the second equality follows from Eq. (\ref{eq:4.12}).

Our aim now is to set a connection between $\Phi$ and $\Psi$
functions at $\lambda=0$ with the classical orbits (\ref{eq:2.12}).
The crucial point is that Eqs. (\ref{eq:4.9}, \ref{eq:4.10}) at $\lambda=0$
go over to
the Jacobi equation (\ref{eq:4.6}) except for the boundary conditions.
But as is known, the Jacobi equation can be obtained by varying the
original Euler-Lagrange equation with respect to the initial data.
Let $\delta z_c$and $\delta{\bar z_c}$ be variations
of Eqs. (\ref{eq:2.12}). Then, the functions
\begin{eqnarray}
u(s)=\delta z_c (\partial^2_{\bar zz}F)_c^{1/2}\exp(+i\int_{0}^{s}Bdt),
\quad
\bar u(s) =\delta\bar z_c (\partial^2_{\bar zz}F)_c^{1/2}
\exp(-i\int_{0}^{s}Bdt)
\label{eq:4.15}\end{eqnarray}
are seen to satisfy the Jacobi equation (\ref{eq:4.6}).

To identify these with the solutions
to Eqs. (\ref{eq:4.9}, \ref{eq:4.10}),
we put
\begin{eqnarray}
\phi(s)&=&(\partial^2_{\bar zz}F)_c^{1/2} \exp(i\int_{0}^{s}Bdt)
\frac{\partial z_c(s)}{ \partial \bar z_F}, \nonumber \\
\bar\phi(s)&=&(\partial^2_{\bar zz}F)_c^{1/2} \exp(-i\int_{0}^{s}Bdt)
\frac{\partial \bar z_c(s)}{ \partial \bar z_F}
\label{eq:4.16}\end{eqnarray}
and similarly,
\begin{eqnarray}
\psi(s) &=&(\partial^2_{\bar zz}F)_c^{1/2} \exp(i\int_{0}^{s}Bdt)
\frac{\partial z_c(s)}{\partial z_I},\nonumber \\
\bar\psi(s)&=&(\partial^2_{\bar zz}F)_c^{1/2} \exp(-i\int_{0}^{s}Bdt)
\frac{\partial\bar z_c(s)}{\partial z_I}.
\label{eq:4.17}\end{eqnarray}
In view of Eq. (\ref{eq:4.14}) we arrive at
\begin{eqnarray}
Det\,\frac{\tilde K}{K_0}&=&\left[\frac{\partial^2_{\bar zz}
F(\bar z(\tau),z(\tau))}{\partial^2_{\bar zz}
F(\bar z(0),z(0))}\right]_c^{1/2}\exp(-i\int_{0}^{\tau}Bdt)
\left(\frac{\partial\bar z_c(0)}{\partial\bar z_F}\right)^{-1}\nonumber\\
&=&\left[\frac{\partial^2_{\bar zz}
F(\bar z(0),z(0))}{\partial^2_{\bar zz}
F(\bar z(\tau),z(\tau))}\right]_c^{1/2}\exp(-i\int_{0}^{\tau}Bdt)
\left(\frac{\partial z_c(\tau)}{\partial z_I}\right)^{-1}.
\label{eq:4.18}\end{eqnarray}

To end the derivation, we compute
$$\frac{\partial^2\Phi_c}{\partial \bar z_F\partial z_I}.$$
It is noteworthy that we take the derivative of the total classical action
including the boundary term. It is the total action $\Phi_c$ that enters the
final answer. A calculation is straightforward and yields
\begin{eqnarray}
\frac{\partial^2\Phi_c}{\partial \bar z_F\partial z_I}
=\frac{1}{2}\left[\partial^2_{\bar zz}F_c(\tau)
\frac{\partial z_c(\tau)}{\partial z_I} +
\partial^2_{\bar zz}F_c(0)
\frac{\partial\bar z_c(0)}{\partial\bar z_F}\right],
\label{eq:4.19}\end{eqnarray}
where we have set $$\partial^2_{\bar zz} F_c(t)\equiv
\partial^2_{\bar zz}F(\bar z(t),z(t))\mid_{c}.$$
Combining Eqs. (\ref{eq:4.18}) and (\ref{eq:4.19}) one finally arrives at
\begin{eqnarray}
\left(Det\,\frac{\tilde K}{K_0}\right)^{-1}=
\frac{1}{[\partial^2_{\bar zz}F_c(\tau)
\partial^2_{\bar zz}F_c(0)]^{1/2}}
\,\frac{\partial^2\Phi_c}{\partial \bar z_F\partial z_I}
\,\exp(i\int_{0}^{\tau}Bdt),
\label{eq:4.20}\end{eqnarray}
which by virtue of (\ref{eq:2.4}) yields
\begin{eqnarray}
{\cal P}^{qc}(\bar z_F,z_I;\tau)=
\exp\left(\Phi_c +\frac{i}{2}\int_{0}^{\tau}Bdt\right)
\left[\frac{1}{[g(\bar z_c(\tau),z_c(\tau))\,g(\bar z_c(0),z_c(0))]^{1/2}}
\,\,\frac{\partial^2\Phi_c}{\partial \bar z_F\partial z_I}\right]^{1/2}.
\label{eq:4.21}\end{eqnarray}

Thus, we have expressed the quasiclassical propagator in terms of the total
classical action and classical orbitals, which is similar to
the DeWitt result for the short time propagator of a particle in a curved
configuration space
(\ref{eq:3.3}). However, there are some important
distinctions. First, the total
action $\Phi$ is involved rather than $S$.
Next, there appears a dependence on the $B$-term. The latter
plays the role of normalization and is necessary to
fix the quantization (by covariant symbols).
This term interpolates between
the covariant and contravariant quantization schemes and disappears
at the point corresponding to the Weyl quantization.
With the aid of the Euler-Lagrange equations
it can also be represented in terms of the extremals:
$$B=\frac{i}{2}(\partial_z\dot z-\partial_{\bar z}\dot{\bar z})\mid_{c}.$$
As has already been mentioned, the total action $\Phi$ and the K\"ahler
potential $F$ are proportional to $l$, whereas it is seen that
$B\sim l^0$, which agrees with suggestion (\ref{eq:qcpr}).

Generalization of the above result to the multidimensional case
(provided one-rank orbits are still considered!) is straightforward.
For instance, consider the compact degenerate $U(N)$ orbit, complex
projective space $CP^{N-1}= U(N)/U(N-1)\otimes U(1)$.
The complex dimensionality of
the manifold is $N-1$ whereas its ${\it rank}=1$. The resulting
K\"ahler potential
$$F=l\log\left[1+\sum_1^{N-1}\bar z^iz^i\right],$$
where $\{z^i,\,i=1,...,N-1\}$ is a complex vector and a positive integer $l$
is the highest weight specifying a representation.
An example is provided by
a quantum system with the dynamical
$U(N)$ symmetry, e.g., generated by a set of bilinears $a^{\dag}_ia_j,\quad
[a_i,a^{\dag}_j]=\delta_{ij},\quad i,j=1,..N$, with
the quasiclassical parameter being a total number of the field excitations:
$l=\sum_in_i$.

An obvious modification of
Eq. (\ref{eq:4.21}) consists in extending Eqs. (\ref{eq:a}) and (\ref{eq:b})
to include vector indices and reads
\begin{eqnarray}
{\cal P}^{qc}=
\exp\left(\Phi_c +\frac{i}{2}\int_{0}^{\tau}tr Bdt\right)
\left[\frac{1}{[g(\bar z_c(\tau),z_c(\tau))\,g(\bar z_c(0),z_c(0))]^{1/2}}
\,\,\det\left(\frac{\partial^2\Phi_c}{\partial \bar z^i_F\partial z^j_I}
\right)\right]^{1/2},
\end{eqnarray}
where $g(\bar z,z)=\det \partial^2_{\bar z^iz^j}F$.

We conclude this section by the following remark.
It seems that the final result (\ref{eq:4.21}) is crucially based on the
fact whether a path integral (\ref{eq:2.9}) exists as a bona-fide integral.
The common belief is however that (\ref{eq:2.9}) cannot be in general
justified as an integral with respect to a certain measure.
Moreover, even the justification of the existence of
the limit in (\ref{eq:2.8}) is rather a nontrivial problem \cite{ber2},
though in the semiclssical domain $(l\to\infty)$ this limit under certain
restrictions does exist, provided expansion in powers of $1/l$
is first carried out \cite{yaff}. However, this procedure does not in general
lead to a genuine integral with respect to {\it a path} measure.

Frequently, the statement occurs (see, e.g., \cite{ercol}) that
once the continuum expression (\ref{eq:2.9}) is concerned,
only formal calculations are possible. This in turn implies that
the continuum form of the path integral may at most provide some hint
about an actual answer which nevertheless is to be obtained in a
rigorous manner within the time-lattice approximation. In general,
this assertion is true, though some exceptions, e.g., the quasiclassical
approximation considered above, are possible.
The justification of the continuum representation in the quasiclassical
domain $(l\gg 1)$ may be given in two formally distinct but in essence
similar ways. One may observe, for instance, that a scalar product
$\langle z_k|z_{k-1}\rangle$ entering into Eq. (\ref{eq:2.8})
is highly peaked about
$\Delta_k(\epsilon)\equiv z_k-z_{k-1}\sim 0$ as $l$ tends to infinity.
This implies that once the leading term ${\cal P}^{qc}$ is concerned,
only terms linear in $\Delta_k$ are to be left in (\ref{eq:2.8}),
which under some
additional mild restrictions on the Hamiltonian function would eventually
lead to continuum representation (\ref{eq:4.3}),
corrections to the leading term
coming, in particular, from higher powers of $\Delta_k$ \cite{yaff}.

We prefer, however, to start from the formal continuum representation
(\ref{eq:2.9})
which makes sense for continuous differentiable paths. Path integral
(\ref{eq:2.9})
is localized on a Hilbert space of the square integrable
paths: $(z,z)<\infty$ \cite{ber2}, so that in general (and actually, almost
surely) $z(s)$ has no time derivative.
To find a way out,
we represent an arbitrary path in the form $z=z_{cl}(s)+\delta z(s)$ and
expand $\delta z(s)$ in a series over an appropriate basis in $[0,\tau]$.
The trick due to Berezin consists in retaining only a finite
part of the series, thereby dealing with continuous and differentiable
paths at all intermediate steps, with the infinite limit being taken only
at the final stage. This procedure converges for the gaussian path integral
(\ref{eq:4.3}) \cite{ber2}, thereby justifying the above continuous calculus.
This is however the case, provided we are concerned as before with
a calculation of the leading term, ${\cal P}^{qc}$.

\vspace{1cm}
{\Large \bf V. Other quantization schemes}
\vspace{1cm}

So far we have been concerned with a specific quantization
scheme, the quantization by covariant symbols.
The next important quantization scheme to be mentioned here
is the contravariant quantization.

The covariant symbol $H^{cov}$ which we identify with $H^{cl}$
is related to the contravariant one
$H^{ctr}$ by
\cite{ber}
\begin{equation}
H^{cov}(\bar z,z)=\int \exp\{\phi(\bar z,z|\bar\xi,\xi)\}H^{ctr}
(\bar\xi,\xi)d\mu(\xi)\equiv\left(\hat{T}H^{ctr}\right)(\bar z,z),
\label{eq:5.1}\end{equation}
where
\begin{equation}
\phi(\bar z,z|\bar\xi,\xi)=F(\bar\xi,z)+F(\bar z,\xi)-F(\bar
z,z)-F(\bar\xi,\xi)
\label{eq:5.2}\end{equation}
and a constant in Eq. (\ref{eq:2.5}) is fixed by
\begin{equation}
\int e^{-F(\bar\xi,\xi)}d\mu(\xi)=1.
\label{eq:5.3}\end{equation}
If the point $(\bar z,z)$ is fixed, the potentials
$-\phi(\bar z,z|\bar\xi,\xi)$ and $F(\bar\xi,\xi)$ generate the very same
metric and are to be related by a group transformation $g_z$:
\begin{equation}
F(\overline{g_z\xi},g_z\xi)=-\phi(\bar z,z|\bar\xi,\xi).
\label{eq:5.4}
\end{equation}
Invariance of the measure $d\mu$ upon $g_z$
along with the normalization (\ref{eq:5.3})
results in $$1^{cov}=1^{ctr}$$ as it should be.

In principle, the operator $\hat T$, being permutable with a group action,
can be expressed via the corresponding Casimir operators. In the case
under consideration
only the second Casimir operator
\begin{equation}
K_2\equiv \Delta= (\partial^2_{z\bar z}F)^{-1}\partial^2_{z\bar z},
\label{eq:5.5}\end{equation}
the Laplace-Beltrami operator with respect to the metric (\ref{eq:2.4}),
is involved.

As is known, in the flat case
\begin{equation}
\hat T(\Delta)=e^{\Delta},
\end{equation}
whereas for the quantization on a sphere and Lobachevsky plane
$\hat T(\Delta)$ has been evaluated in a form of infinite products
\cite{ber1}.

As is seen from Eq. (\ref{eq:4.21}), both the expression in the brackets
and the $B$-term are of an order of ${\cal O}(1)$ whereas $\Phi_c={\cal
O}(l)$.  This means that $H^{ctr}$ is to be taken with the accuracy up to
an order of ${\cal O}(1)$. Therefore, for our purposes we need an asymptotic
relation between symbols rather than the exact one, which is not easily
available.  This is pursued by the following notice. Being nonpositive, the
function $\phi$ reaches zero at the point $(\bar\xi,\xi)=(\bar z,z)$.  As $l$
goes to infinity, the maximum becomes sharper localizing $\phi$ at
$(\bar\xi,\xi)=(\bar z,z)$.  Developing then the integrand
in (\ref{eq:5.1}) in
powers of $\eta=\xi-z,\quad \bar\eta=\bar\xi-\bar z$ one gets
\begin{eqnarray}
H^{cov}(\bar z,z)&=&\int e^{-{\bar\eta}\eta}\left[H^{ctr}(\bar z,z)+
+\Delta H^{ctr}(\bar z,z){{\bar\eta}\eta}\right]
\frac{d\bar\eta d\eta}{2\pi i}
+o(1)\nonumber \\
&=&[1+\Delta+{\cal O}(1/l^2)]\,H^{ctr}(\bar z,z), \quad
l\to \infty
\label{eq:5.6}\end{eqnarray}
In view of this, one may convert Eq. (\ref{eq:4.21}) in a form suitable
for the quantization by contravariant symbols.

To conclude this section, we will specify Eq. (\ref{eq:4.21}) for the flat
$(M=C)$ case relevant for the Heisenberg-Weyl coherent states. To avoid
confusion with dimensions, we introduce, following Ref. \cite{tuynman},
coordinates
$x=p/\alpha$ and $y=q/\beta$ and the complex
dimensionless coordinate
$z=1/\sqrt{2}\,(x+iy)$. Constants $\alpha$ and $\beta$ are of dimensions
of momentum and position, respectively. It is convenient to introduce
the dimensionless constant
$$
\gamma=\alpha\beta/\hbar$$ which plays the role of the representation
index $l$. The classical limit becomes quite transparent in this notation.
It means a passage from systems of units to measure $\alpha$ and $\beta$
quite adequate for quantum description to those that are more convenient
for the classical one. For instance, if one chooses a "classical scale"
$\alpha=1m,\beta=
1kg\, m/sec$, then ${\gamma}^{-1}\approx 10^{-34}$,
which effectively corresponds
to small $\hbar$. It is just in this sense that one should understand
the limit $\hbar\to 0$.

The conventional $2$-form $w=dp\wedge dq$ goes over to
$$w=- \alpha\beta\, dx\wedge dy,$$ so that
$$w/\hbar=i\gamma \,d\bar z\wedge dz.$$ We introduce a set of the $\gamma$-
dependent Hesenberg-Weyl coherent states:
\begin{eqnarray}
|z;\gamma\rangle=\exp(-\frac{\gamma}{2}\bar zz+\sqrt{\gamma}za^{\dag})\,
|0\rangle,
\label{eq:5.7}\end{eqnarray}
whence
\begin{eqnarray}
F=\log |\langle 0|z;\gamma\rangle|^{-2}=\gamma\bar zz.
\label{eq:5.8}\end{eqnarray}

In the flat case covariant symbols are related to those in the
$\alpha$-quantization scheme by
$$H^{cov}(\bar z,z)=(\hat T_{\alpha}(\Delta)H^{(\alpha)})(\bar z,z),
\quad \hat T_{\alpha}(\Delta) =e^{\alpha\Delta}, \quad \alpha\in [0,1],.$$
the covariant, contravariant and Weyl quantization schemes being specified
by $\alpha=0,1$ and $1/2$, respectively.
As a result one gets
$$H^{cov}-\frac{1}{2}\Delta H^{cov}\equiv
H^{cov}-[(\frac{1}{2}-\alpha)+\alpha]\Delta H^{cov}
=H^{(\alpha)}-(\frac{1}{2}-\alpha)\Delta H^{cov}+{\cal O}(1/l^2),$$
which yields
\begin{eqnarray}
{\cal P}_{flat}^{(qc)}
&=&\left[\frac{1}{\gamma}\frac{\partial^2\Phi_c}
{\partial \bar z_F\partial z_I}
\right]^{1/2}\exp\left\{\Phi_c+\frac{i}{2}
\int\limits_{0}^{\tau}Bdt\right\}\nonumber\\
&=&\left[\frac{1}{\gamma}\frac{\partial^2\Phi_c}
{\partial \bar z_F\partial z_I}
\right]^{1/2}\exp\left\{\Phi^{(\alpha)}_c+i(\frac{1}{2}-\alpha)
\int\limits_{0}^{\tau}Bdt\right\}
\label{eq:5.9}\end{eqnarray}
\begin{equation}
\sim\left[\frac{1}{\gamma}\frac{\partial^2\Phi^{(\alpha)}_c}
{\partial \bar z_F\partial z_I}
\right]^{1/2}\exp\left\{\Phi^{(\alpha)}_c+i(\frac{1}{2}-\alpha)
\int\limits_{0}^{\tau}B^{(\alpha)}dt\right\},
\label{eq:5.91}\end{equation}
where equivalence classes are defined by
$f\sim g =\{\,f\,|\,f/g=1+o(1),\quad l\to\infty\},$ so that $H^{(\alpha)}\sim
H^{cov}\equiv H^{cl}.$
This result (the first line in Eq. (\ref{eq:5.9})), with the $B$-term
however being missed,
was derived by
Weissman \cite{weiss} by extending Miller's semiclassical algebra to the
coherent-state setting.\footnote{ In a subsequent paper \cite{weiss2}
the author, to recover the correct result,
was forced to allow for the $B$-term in a specific case
of the parametric amplifier}
Originally, Miller's formalism incorporated
eigenstates of Hermitian operators to relate a quantum mechanical matrix
element of a general unitary trasformation, in the semiclassical limit,
to a generator of a corresponding canonical transformation \cite{miller}.

It is to be noted that in deriving Eqs. (\ref{eq:5.9},\ref{eq:5.91})
original equations of motion (\ref{eq:2.12})
that correspond to the covariant quantization have been kept fixed.
That is why (\ref{eq:5.91}) cannot be regarded as a genuine
$\alpha$-representation. To derive the latter, one would have to start with
equation $\delta\Phi^{(\alpha\neq 0)}=0$, whose solutions in contrast to
(\ref{eq:2.12}) would be bearing an explicit $l$-dependence, namely,
$z^{(\alpha)}=z^{cl}+{\cal O}(1/l).$ In that case, however, it would be
natural to start, instead of Eq. (\ref{eq:2.7}), with the $\alpha$-symbol
of the evolution operator.

\vspace{1cm}
{\Large \bf VI. Examples}
\vspace{1cm}

In this section, we will illustrate the basic result (\ref{eq:4.21})
with a few simple but instructive examples.

{\it $SU(2)$ path integral ($M=CP^1$, compact manifold)}\\
Coherent state  for the UIR' of the $SU(2)$ group is given by
\begin{equation}
|z;j\rangle=(1+|z|^2)^{-j}\exp (zJ_{+})\,|j;-j\rangle \,,
\label{eq:6.1}
\end{equation}
where $z\in SU(2)/S(U_1\times U_1)\simeq CP^1$, which can be thought of
as an extended complex plane $\bar C^1$.
The operators $J_{\pm},J_0$ span the $SU(2)$ algebra
$$[J_0,J_{\pm}]={\pm} J_{\pm}\,,\,\,\,\,\,\, [J_{+},J_{-}]=2J_0\,$$
and the lowest weight state $|j;-j\rangle$ is annihilated by $J_{-}$. From
Eq. (\ref{eq:2.2}) it follows that
$$F(\bar z,z)=2j\log(1+\bar zz), \quad l=2j\in N,$$
where $j$ must be
a half-integer corresponding to the unitary irreducible representations
of $SU(2)$. From the geometric viewpoint, this requirement is
to be imposed in order that a holomorphic prequantum line bundle over $CP^1$
can be constructed.
The general Eq. (\ref{eq:2.9}) reads ($\hbar=1$)
\begin{eqnarray}
&&{\cal P}_j(\bar z_F,z_I;\tau)=\int_{z(0)=z_I}^{{\bar z}(\tau)
={\bar z}_F}D\mu(z)\frac
{(1+{\bar z}_Fz(\tau))^j
(1+{\bar z}(0)z_I)^j}
{(1+|z_F|^2)^j(1+|z_I|^2)^j} \times  \nonumber \\
&& \exp\left(j\int_{0}^{\tau} \frac{{\dot{\bar z}}(s)z(s)
-{\bar z(s)}\dot z(s)}
{1+\bar z(s)z(s)}ds -i\int_{0}^{\tau}H^{cl}({\bar z(s)},z(s))ds\right),
\label{eq:6.2}
\end{eqnarray}
whith the normalization
${\cal P}_j{\mid}_{H=0}=\langle z_F;j| z_I;j\rangle.$
Here $D\mu_j(z)$ stands for the infinite pointwise product of
the $SU(2)$ invariant measures
$$d\mu_j=\frac{2j+1}{2\pi i}\frac{dzd\bar z}{(1+|z|^2)^2}.$$

As a simple but nontrivial example that directly demonstrates
the usefulness of Eq. (\ref{eq:4.21}) consider a
system governed by the Hamiltonian \cite{koch3}
$$H=2A(t)J_z+f(t)J_{+}+\bar f(t)J_{-}.$$
The stationary-phase equations read
\begin{eqnarray}
i\dot z=2A(t)z+f(t)-\bar f(t)z^2,\quad z(0)=z_I,
\label{eq:6.3}\end{eqnarray}
\begin{eqnarray}
-i\dot{\bar z}=2A(t)\bar z+\bar f(t)-f(t)\bar z^2,
\quad \bar z(\tau)=\bar z_F.
\label{eq:6.4}\end{eqnarray}
Being of the Riccati type these equations cannot be solved explicitly, but
yet some important information is available. Namely, let us look for the
solutions to Eqs. (\ref{eq:6.3},\ref{eq:6.4}) in the form
\begin{eqnarray}
z(t)=\frac{a(t)z_I+b(t)}{-\bar b(t)z_I+\bar a(t)},\quad a(0)=1,b(0)=0,
\quad {\left(\begin{array}{cc}
a & b \\
{-\bar b} & {\bar a}
\end{array}\right)\in SU(2)},
\label{eq:6.5}\end{eqnarray}
and
\begin{eqnarray}
\bar z(t)=\frac{\bar d(t)\bar z_F+\bar c(t)}{-c(t)\bar z_F+d(t)},\quad
d(\tau)=1,c(\tau)=0,
\quad {\left(\begin{array}{cc}
{\bar d} & {\bar c} \\
-c & d
\end{array}\right)\in SU(2)},
\label{eq:6.6}\end{eqnarray}
respectively.
In view of Eq. (\ref{eq:6.3}) the functions $a(t)$ and $b(t)$ satisfy
\begin{eqnarray}
\dot a&=&-iAa+if\bar b\nonumber\\
\dot b&=&-iAb-if\bar a,
\label{eq:6.7}\end{eqnarray}
whereas the functions $d(t)$ and $c(t)$,
due to Eq. (\ref{eq:6.4}),
are to be taken to satisfy
\begin{eqnarray}
\dot d=-iAd+if\bar c\quad \dot c=-iAc-if\bar d.
\label{eq:6.8}\end{eqnarray}
The solutions to these equations are related to those of Eqs. (\ref{eq:6.7})
by
$$\bar d(t)=\bar a(t)a(\tau)+\bar b(t)b(\tau),\quad
c(t)=-a(t)b(\tau)+b(t)a(\tau).$$

Despite the fact that we do not have any explicit solutions
to Eqs. (\ref{eq:6.3},\ref{eq:6.4}) we have managed to explicitly
determine their dependence on initial data, $z_I$ and $\bar z_F$.
This, in fact, is all we need to apply Eq. (\ref{eq:4.21}).
After some algebra one gets
$$\Phi_c =2j\log
[{\bar a}(\tau)-{\bar b}(\tau)z_I+b(\tau)\bar z_F+
a(\tau)\bar z_Fz_I] -j\log(1+|z_F|^2)(1+|z_I|^2),$$
$$B=(2A-f\bar z-\bar fz)\mid_c=-i\frac{d}{dt}\log
\frac{-\bar b(t)z_I+\bar a(t)}{-c(t)\bar z_F+d(t)},$$
$$\frac{\partial^2\Phi_c}{\partial \bar z_F\partial z_I}=
\frac{2j}
{[{\bar a}(\tau)-{\bar b}(\tau)z_I+b(\tau)\bar z_F+
a(\tau)\bar z_Fz_I]^2}.$$
By plugging all this into Eq. (\ref{eq:4.21}) one finally arrives at
$${\cal P}_j(\bar z_F,z_I;\tau)= \exp\,\Phi_c,$$
which coincides with direct time-lattice calculations.
This result agrees with the Diustermaat-Heckman
theorem which states that
the WKB expansion is exact, provided a Hamiltonian phase flow
leaves a metric tensor invariant (\ref{eq:lie}). This is just the case
for $$H=2A(t)J_z+f(t)J_{+}+\bar f(t)J_{-}.$$
Moreover, the dynamical invariance,
i. e., the fact that $H$ belongs to the $SU(2)$ algebra, results in
$${\cal P}_{red}=1,$$ which is of importance in deriving
the generalized Bohr-Sommerfeld quantization conditions \cite{koch3}.

{\it Path integral for the Heisenberg-Weyl coherent states
($M=C$, noncompact manifold)}\\
By virtue of Eq. (\ref{eq:5.8}), the general representation (\ref{eq:2.9})
reduces to
\begin{eqnarray}
&&{\cal P}_{\gamma}=
\langle z_F,\gamma|\,\exp-i\int\limits_{0}^{\tau}Hds\,|z_I,\gamma\rangle
= \int\limits_{z(0)=z_I}^{\bar z(\tau)=\bar z_F}DzD\bar z\,
\exp\left\{\frac{\gamma}{2}\int\limits_{0}^
{\tau}(z{\dot{\bar z}}-\bar z\dot z)ds
\right.\nonumber \\
&&-i\int\limits_{0}^{\tau} H^{cl}(\bar z,z)ds
\left.+\frac{\gamma}{2}[\bar z_Fz(\tau)+\bar z(0)z_I-|z_F|^2-|z_I|^2]\right\}
\label{eq:6.9}
\end{eqnarray}
with the normalization
${\cal P}_{\gamma}({\bar z}_F, z_I;\tau) \mid_{H=0}
=\,\,\langle z_F,\gamma|z_I,\gamma\rangle\,.$

For a harmonic oscillator ($\hbar=1$) $H=\omega a^{\dag}a$
one gets $H^{cl}=H^{(\alpha=0)}
=\gamma\omega |z|^2,\quad H^{(\alpha)}=H^{cl}-\alpha\omega$
and solutions to
Eqs. (\ref{eq:2.12}) read
\begin{equation} z_c(s)=z_I\exp(-i\omega s)\quad
\bar z_c(s)=\bar z_F \exp(-i\omega(\tau-s)), \label{eq:6.10}\end{equation}
which
in turn results in $$ \Phi^{(\alpha)}_c=\gamma\bar z_Fz_I
\exp(-i\omega\tau)-\frac{\gamma}{2}(|z_F|^2 +|z_I|^2)+i\tau\alpha\omega,
\quad B=\omega.$$
Equation (\ref{eq:5.9}) can be applied to yield
$${\cal P}_{\gamma}=
\exp\Phi^{(\alpha=0)}_c,$$ as it should be.

For gaussian actions the path integral (\ref{eq:6.9}) reduces to
Eq. (\ref{eq:5.9}). However, in the case when Hamiltonian cannot be cast
into a linear combination of
the oscillator group generators $a^{\dag}a,\, a^{\dag}$ and $a$, the
quasiclasscal propagator (\ref{eq:5.9}) does not merely reduce
to the simple form $$\exp\Phi_c.$$  For instance, for the Hamiltonian
of a parametric amplifier
$$H=\omega a^{\dag}a-\frac{g}{2}[{a^{\dag}}^2e^{-2i\omega t}+
a^2e^{2i\omega t}]$$
one gets ($\gamma=1$)
$$z_c(s)\exp(i\omega s)=
\frac{i\bar z_F-z_I\sinh g\tau}{\cosh g\tau}\sinh gs
+z_I\cosh gs$$
$$\bar z_c(s)\exp(i\omega (\tau-s))=
\frac{\bar z_F+iz_I\sinh g\tau}{\cosh g\tau}\cosh gs
-iz_I\sinh gs$$
$$\Phi_c=\frac{\bar z_F z_Ie^{-i\omega\tau}}{\cosh g\tau}
+\frac{i}{2}\tanh g\tau (\bar z^2_F
+z^2_I)-\frac{1}{2}(|z_F|^2+|z_I|^2),\quad B=\omega.$$
Equation (\ref{eq:5.9}) is again exact and reads
$${\cal P}(\bar z_F,z_I;\tau)= (\cosh g\tau)^{-1/2} \exp\Phi_c,$$
which coincides with the direct time-lattice calculations \cite{hillery}.
Equation (\ref{eq:4.21}) is thus seen to correctly recover
the known results for the exact propagators.

\vspace{1cm}
{\Large \bf VII. Conclusion}
\vspace{1cm}

In the present paper, we have discussed quasiclassical
quantization of a classical mechanics on a symplectic group orbit
of ${\it rank}=1$ in terms of the relevant
coherent-state path integral, which provides an appropriate
expansion of quantum-mechanical propagator for large values of the
highest weight $l$ that specifies the underlying group representation.
The principal result is a new explicit quasiclassical formula
for a propagator on a coherent-state manifold
which is written entirely in terms of classical data
and reveals the leading large $l$ behavior of the propagator.
This representation is important since a wealth of physically relevant
classical phase spaces
admit a natural K\"ahler polarization, e.g., $S^2\simeq CP^1$,
the classical phase space for a spin or $S^{1,1}\cong D^1$, a unit disk
on a complex plane---a natural phase space for models
of quantum optics \cite{inom}.
In this regard, the quasiclassical representation
(\ref{eq:4.21}) can be applied to study, e.g., spin tunneling in the
semiclassical limit \cite{enz} and related problems \cite{kurch} as well as
the behavior of highly excited field states in quantum optical models.

As for possible generalizations of
(\ref{eq:4.21}), one might proceed in three directions.
First, it would be desirable to extend the result to higher-rank
phase spaces. Second, it would be of practical importance
to generalize (\ref{eq:4.21}) to non-local actions that arise
provided certain degrees of freedom
in an original Hamiltonian can be integrated out. This is the case
for a large class of interactions that involve
bilinear combinations of Lie algebra generators and field coordinates.
Third, it would be interesting to extend this approach to the supersetting,
since even the simplest super phase spaces
happen to be relevant to important physics.
For instance, one-rank degenerate orbit of the $SU(2|1)$ supergroup
can be viewed as a phase space of the $t-J$ model of strongly correlated
electrons which is believed to adequately describe a
high $T_c$ superconducting state.

\vspace{1cm}
{\Large \bf Acknowledgments}
\vspace{1cm}

This research was supported by Russian
Foundation for Fundamental Research under Grant No 96-01-00223.
It is a pleasure for me to thank Professor Hajo Leschke for his hospitality
at the University Erlangen-N\"urnberg, Erlangen, Germany, where a final part
of this work was carried out.

%\newpage

\end{document}